\documentclass[twocolumn,tighten,times]{aastex62}

\hypersetup{linkcolor=blue,citecolor=blue,filecolor=cyan,urlcolor=magenta}

\newcommand{\ratio}{{\langle B_{t}^{2} \rangle}/{\langle B_{0}^{2} \rangle}}

\newcommand{\um}{$\mu$m}

\shorttitle{The strength and morphology of the magnetic field in the galactic outflow of M82}
\shortauthors{Lopez-Rodriguez et al.}

\begin{document}

\title{The strength and structure of the magnetic field in the galactic outflow of M82}

\correspondingauthor{Lopez-Rodriguez, E.}
\email{elopezrodriguez@stanford.edu}

\author{Enrique Lopez-Rodriguez}
\affil{Kavli Institute for Particle Astrophysics and Cosmology (KIPAC), Stanford University, Stanford, CA 94305, USA}
\affil{SOFIA Science Center, NASA Ames Research Center, Moffett Field, CA 94035, USA}

\author{Jordan A. Guerra}
\affil{Department of Physics, Villanova University, 800 E. Lancaster Ave., Villanova, PA 19085, USA}

\author{Mahboubeh Asgari-Targhi}
\affil{Harvard-Smithsonian Center for Astrophysics, 60 Garden Street MS-15, Cambridge, MA 02138, USA}

\author{Joan T. Schmelz}
\affil{SOFIA Science Center, NASA Ames Research Center, Moffett Field, CA 94035, USA}
\affil{Universities Space Research Association (USRA), 7178 Columbia Gateway Drive, Columbia, MD 21046, USA}



\begin{abstract} 
Galactic outflows driven by starbursts can modify the galactic magnetic fields and drive them away from the galactic planes. Here, we quantify how these fields may magnetize the intergalactic medium. We estimate the strength and structure of the fields in the starburst galaxy M82 using thermal polarized emission observations from SOFIA/HAWC+ and a potential field extrapolation commonly used in solar physics. We modified the Davis-Chandrasekhar-Fermi method to account for the large-scale flow and the turbulent field. Results show that the observed magnetic fields arise from the combination of a large-scale ordered potential field associated with the outflow and a small-scale turbulent field associated with bow-shock-like features. Within the central $900$ pc radius, the large-scale field accounts for $53\pm4$\% of the observed turbulent magnetic energy with a median field strength of $305\pm15$ $\mu$G, while small-scale turbulent magnetic fields account for the remaining $40\pm5$\% with a median field strength of $222\pm19$ $\mu$G. We estimate that the turbulent kinetic and turbulent magnetic energies are in close equipartition up to $\sim2$ kpc (measured), while the turbulent kinetic energy dominates at $\sim7$ kpc (extrapolated). We conclude that the fields are frozen into the ionized outflowing medium and driven away kinetically. The magnetic field lines in the galactic wind of M82 are `open,'  providing a direct channel between the starburst core and the intergalactic medium. Our novel approach offers the tools needed to quantify the effects of outflows on galactic magnetic fields as well as their influence on the intergalactic medium and evolution of energetic particles.
\end{abstract} 



\section{Introduction} \label{sec:int}

Messier 82 (M82) is a canonical starburst galaxy at a distance of $3.85\pm0.35$ Mpc  \citep[20 pc/arcsec,][using SNIa 2014J]{Vacca2015}. Observations reveal a bipolar superwind that originates in the core and extends perpendicular to the galactic plane out into the halo and intergalactic medium (IGM) \citep[e.g.][] {SBH1998,Lehnert1999,Ohyama2002,Engelbracht2006,Heckman2017}. $H_{\alpha}$ emission line reveals an extended and continuous structure up to $\sim11$ kpc with an emission line structure in the northern region at $\sim11-12$ kpc, which is known as the `cap' \citep{DB1999}. The $H_{\alpha}$ is thought to be excited by radiation from the starburst region along the superwind. X-ray emission is observed at the location of the starburst region but also spatially coincident with the northern cap \citep{Lehnert1999}. X-ray emission is thought to be generated due to shock heating driven by the collisions between the superwind and massive ionized clouds in the halo of M82. The $H_{\alpha}$-X-ray spatial correlation show evidence of this superwind driven material out of the galactic plane of M82. 

The geometry of the magnetic fields in the core of the galaxy and the superwind has been investigated with various observing techniques. Non-thermal radio emission from the central region extends normal to the plane of M82, suggesting that the synchrotron emitting plasma is part of the outflow \citep{Reuter1992}. These results were obtained with the Very Large Array (VLA) from 3.6 cm to 90 cm at an angular resolution in the range of $4-35$\arcsec. Subsequent observations with the VLA at 3.6 and 6.2 cm at a resolution of $15$\arcsec~found evidence for a poloidal magnetic field at heights of up to 400 pc from the plane, consistent with an outflowing plasma with velocities high enough to drag the field along with it \citep{Reuter1994}. These data from the VLA archive were later combined with complementary 18 cm and 22 cm data at a resolution of $15$\arcsec~from the Westerbork Synthesis Radio Telescope (WSRT) by \citet{Adebahr2017}. The analysis reveals polarized emission with a planar geometry in the inner part of the galaxy, which they interpret as a magnetized bar. The longer wavelength data show polarized emission up to a distance of 2 kpc from the disk, which could be described as large-scale magnetic loops in the halo.

Results from optical and near-infrared interstellar polarization observations suggest that the field geometry is perpendicular to the plane of this edge-on galaxy, in line with the superwind, rather than parallel as might be expected for spiral galaxies \citep[see e.g.][and references therein]{Jones2000}. However, polarimetry at optical and near-infrared wavelengths is strongly contaminated by the scattering of light from the nuclear starburst, and the relativistic electrons that give rise to the radio synchrotron emission may not sample the same volume of gas as polarization. Fortunately, the magnetic field structure can also be obtained using observations of polarized thermal emission from the aligned dust grains in the central region of M82. The field lines in the 850 \um~polarization map with an angular resolution of $15$\arcsec~from the Submillimetre Common-User Bolometer Array (SCUBA) camera on the James Clerk Maxwell Telescope (JCMT) formed a giant magnetic loop or bubble with a diameter of at least 1 kpc. \citet{Greaves2000} speculated that this bubble was possibly blown out by the superwind. However,  a map created from reprocessed data with an angular resolution of $20$\arcsec~did not show a clear bubble \citep{Matthews2009}.

\citet{Jones2019} analyzed far-infrared polarimetry observations of M82 at $53$ and $154$ \um~with angular resolutions of $4.85$\arcsec~and $13.6$\arcsec, respectively, taken with the High-resolution Airborne Wideband Camera-plus \citep[HAWC+;][]{Vaillancourt2007,Dowell2010,Harper2018} on the Stratospheric Observatory for Infrared Astronomy (SOFIA). The polarization data at both wavelengths reveal a magnetic field geometry that is perpendicular to the disk near the core and extends up to 350 pc above and below the galactic plane. The 154 \um~data add a component that is more parallel to the disk further from the nucleus. These results are consistent with the interpretation that the superwind outflow is dragging the field along with it.

The somewhat controversial nature of the magnetic bubble detected by \citet{Greaves2000} but not by \citet{Matthews2009} using the same data inspired us to draw upon a solar physics analogy. Would the HAWC$+$ field lines described by \citet{Jones2019} extend towards the IGM (galactic outflow), like the magnetic environment in the solar wind, or turn over to the galactic plane (galactic fountain), similar to coronal loops? To extend the HAWC$+$ data to greater heights above and below the galactic plane, we turn to a standard and well-tested technique used in heliophysics $-$ the potential field extrapolation. With only rare exceptions \citep[see e.g.][ and references therein]{Schmelz1994}, the magnetic field in the solar corona, a magnetically dominant environment, cannot be measured directly. Therefore, significant effort has been invested by the community into extrapolating the field measured at the surface via the Zeeman Effect up into the solar atmosphere. The simplest of these approximations assumes that the electrical currents are negligible so the magnetic field has a scalar potential that satisfies Laplace equation and two boundary conditions: it reduces to zero at infinity and generates the normal field measured at the photosphere. The pioneering work by \citet{Schmidt1964} assumed a flat photospheric boundary. \citet{Sakurai1982} later expanded this technique to include a spherical boundary surface in a code that was available and widely used.

In this paper, we have quantified the magnetic field strength and structure in the starburst region of M82. We have also modified the solar potential field approximation to work with the HAWC$+$ data in order to extrapolate the magnetic field observed by \citet{Jones2019} and investigate the magnetic structures in the halo of M82. The paper is organized as follows: Section \ref{sec:Bfield} describes the estimation of the averaged magnetic field strength in the starburst region, which we use to compute a two-dimensional map of the energies in Section \ref{sec:2dB}. The potential field extrapolation is developed in Section \ref{sec:PF}. We discuss the results in Section \ref{sec:DIS} and conclusions in Section \ref{sec:CON}.


\section{The Magnetic Field Strength of M82}\label{sec:Bfield}

\subsection{The Davis-Chandrasekhar-Fermi Method}\label{subsec:Bclassic}

The plane-of-the-sky (POS) magnetic field strength has been estimated from  polarimetric data in Galactic \citep[i.e.][Li et al., in prep.]{Wentzel1963,Schmidt1970,Gonatas1990,Zweibel1990,Leach1991,Shapiro1992,Morris1992,Chrysostomou1994,Minchin1994,Aitken1998,Davis2000,Henning2001,Attard2009,Cortes2010,Chapman2011,Stephens2013,Cashman2014,Zielinski2020} and extragalactic sources \citep[i.e.][]{Lopez-Rodriguez2013,Lopez-Rodriguez2015,Lopez-Rodriguez2020} by using the classical Davis-Chandrasekhar-Fermi (DCF) method \citep{Davis1951,CF1953}. This method relates the line-of-sight velocity dispersion and the plane-of-sky polarization angle dispersion. It assumes an isotropically turbulent medium, whose turbulent kinetic and turbulent magnetic energy components are in equipartition. 

For a steady state with no large-scale flows, the DCF method establishes that the velocity of a transverse magnetohydrodynamical (MHD; Alfv\'en) wave, 

\begin{equation}\label{eq:Va}
    V_{A} = \frac{B}{\sqrt{4\pi\rho}},
    \label{eq:alfven_v}
\end{equation}

\noindent
is related to the observed dispersion of polarization angles where $B$ is the magnetic field strength and $\rho$ is the mass density. This relation is derived from the wave equation with the propagation velocity found to be

\begin{equation}\label{eq:V2}
    V^{2} = V_{A}^{2} = \frac{\sigma_{v}^{2}}{\sigma_{\phi}^{2}},
    \label{eq:wave_v}
\end{equation}

\noindent
where $\sigma_{v}$ is the amplitude of the time variation ({\it i.e.}, velocity dispersion) and $\sigma_{\phi}$ is the spatial amplitude (i.e. angular dispersion). Combining Eqs. \ref{eq:alfven_v} and \ref{eq:wave_v}, gives us the well-known DCF approximation:

\begin{equation}
    B_{\rm DCF} = \xi\sqrt{4\pi\rho}\frac{\sigma_{v}}{\sigma_{\phi}}.
    \label{eq:DCF}
\end{equation}

\noindent
The angular dispersion $\sigma_{\phi}$ is the standard deviation of the distribution of polarization angles. When $\sigma_{\phi} \le 25^{\circ}$, $\xi \sim 0.5$ accounts for the projection of the magnetic field and density distributions on the POS \citep{Ostriker2001}.

\begin{figure}[ht!]
\includegraphics[angle=0,scale=0.31]{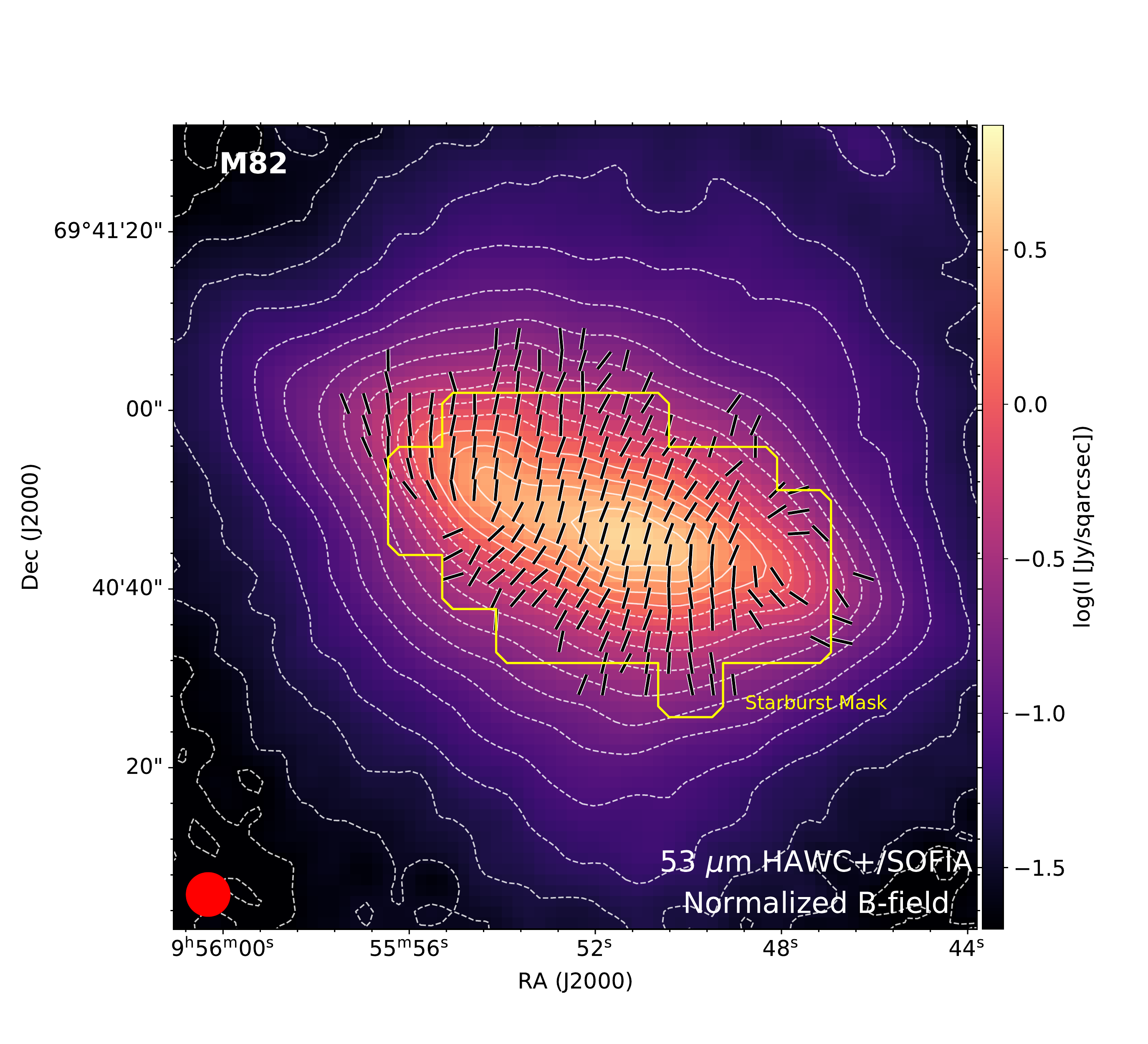}
\caption{Inferred magnetic field orientation (black) of M82 using the 53 \um~polarimetric observations with HAWC+/SOFIA \citep{Jones2019}. All polarization measurements are normalized to unity, and only those with $p/\sigma_{p} \ge 2$ are shown. The color scale shows the total intensity at 53 \um. Contours start at $4\sigma$ with increments of $2^{n\sigma}$, where $n = 2.0,2.5, 3.5, \dots$ and $\sigma = 2.52$ mJy/sqarcsec. The starburst mask (yellow solid line), with a size of $873\times 510$ pc$^{2}$ at a PA of $\sim69^{\circ}$, taken from \citet[][fig. 18]{Contursi2013} is used for the parameter estimations of the magnetic field strengths. Beam size (red circle) is shown.
\label{fig:fig1}}
\epsscale{2.}
\end{figure}

Figure \ref{fig:fig1} shows the magnetic field orientations of M82 inferred from the 53 \um~observations with HAWC+ \citep{Jones2019}. Polarization measurements with $P/\sigma_{P} \ge 2$ are shown, where $\sigma_{P}$ is the uncertainty of the polarization fraction, $P$. Polarization measurements have been normalized and rotated by $90^{\circ}$ to show the B-field orientation. To apply the DCF approximation, the velocity dispersion, mass density, and angular dispersion have to be estimated in the same region. We perform our DCF analysis within the starburst region generated using the gas kinematics in the superwind \citep[fig. 18 from][]{Contursi2013}. Specifically, the FIR spectroscopic analysis of \citet{Contursi2013} separated the kinematics of M82 into four regions $-$ the disk, the starburst, the northern outflow, and the southern outflow $-$ based on several emission lines ([CII], [OI] and [OIII]) using PACS/\textit{Herschel} data. We find that only the starburst region provides sufficient polarization measurements to perform the dispersion analysis used for this work. The starburst mask, with a size of $873\times510$ pc$^{2}$ at a position angle (PA) of $\sim 69^{\circ}$ in the east of north direction,  the region of focus for this work, is shown in Figure \ref{fig:fig1}.

\begin{figure*}[ht!]
\includegraphics[angle=0,scale=0.54]{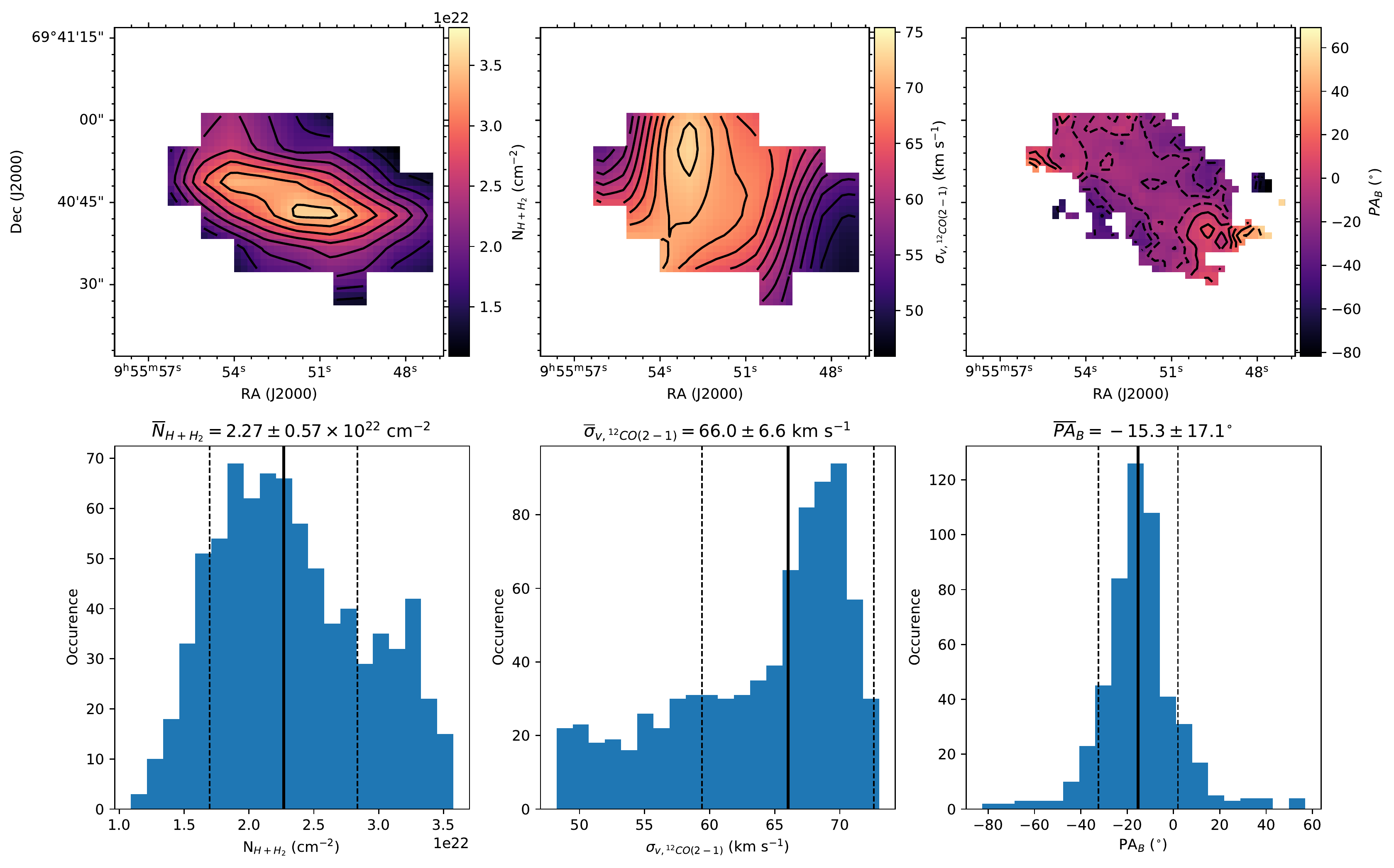}
\caption{\textit{Top:} Maps of the $N_{H+H_{2}}$ (left) from \citet{Jones2019}, $\sigma_{v,^{12}\rm CO(1-2)}$ (middle) from \citet{Leroy2015}, and PA of the inferred B-field orientation, $PA_{B}$, (right) using the 53~\um~polarimetric observations from \citet{Jones2019}. Contours start at $N_{H+H_{2}} = 1.09 \times 10^{22}$ cm$^{-2}$ and $\sigma_{v,^{12}\rm CO(1-2)} = 48$ km s$^{-1}$, and increase in steps of $0.3 \times 10^{22}$ cm$^{-2}$, and $2$ km s$^{-1}$, respectively. The contours of $PA_{B}$ are displayed in the range of $[-90,90]^{\circ}$ in intervals of $10^{\circ}$. \textit{Bottom:} Histograms of the measurements for each map. The median (solid line) and $1\sigma$ uncertainties (dashed lines) are shown.}
\label{fig:fig2}
\epsscale{2.}
\end{figure*}

Figure \ref{fig:fig2}-top shows the maps of the several empirical quantities necessary to estimate the magnetic field strength within the starburst mask. Figure \ref{fig:fig2}-bottom shows the distributions of each quantity with the estimated median (solid line) and $1\sigma$ uncertainties (dashed lines). The first is the column density, $N_{H+H_{2}}$, from \citet{Jones2019}.  The $N_{H+H_{2}}$ map has been smoothed using a Gaussian profile with a full-width-at-half-maximum (FWHM) equal to the beam size of the HAWC+ observations and projected to the HAWC+ observations. We estimate the median column density within the starburst mask to be $N_{H + H_{2}} = (2.27\pm0.57)\times10^{22}$ cm$^{-2}$. To convert $N_{H + H_{2}}$ to mass density ($\rho$), we need to know the extend of the gas and dust in the line-of-sight (LOS) direction. An estimate of this dimension is the effective depth of the starburst region, $\Delta^{\prime}$, which can be calculated following \cite{Houde2011}. Specifically, we calculate the normalized autocorrelation function of the polarized flux of M82 using polarization measurements with $P/\sigma_{P} \ge 3$. This cut ensures that the same polarization measurements are used through the data analysis. Then, the half-width-at-half-maximum (HWHM) of the distribution is taken as the value of $\Delta^{\prime}$, which is estimated to be $\Delta' = 0\farcm1613$ ($193.6$ pc). We interpret $\Delta^{\prime}$ as the depth of the starburst region that contains $\ge50$\% of the polarized flux, which assumes that the gas and dust distribution in the starburst of M82 is isotropic. Note that this estimation does not use a filling factor of the dust within the starburst region. Our result is in good agreement with that of \citet{Adebahr2017}, who estimated a width of the polarized emitting region of $\sim 250$ pc using $3$ and $6$ cm data and assuming a cylindrical symmetry. Finally, using $\Delta^{\prime}$, we can estimate the mass density within the starburst mask to be $\rho = m_{H}\mu N_{H + H_{2}}/\Delta' = (1.78\pm0.45)\times10^{-22}$ g cm$^{-3}$, where $m_{H}$ is the hydrogen mass and $\mu = 2.8$ is the mean molecular mass.

Multiple authors have measured the velocity dispersion for the central disk of M82 within similar physical scales as our mask. Emission lines, e.g. [ArIII] 8.99 \um, H$_{2}$ 17.034 \um, [FeII] 25.98 \um, and [SiII] 34.815 \um, in the mid-infrared observed with the Infrared Space Observatory (ISO) and an angular resolution of $9\arcsec$ (180 pc) show velocities within a range of 71$-$114 km s$^{-1}$ \citep{Sohn2001}. $^{12}$CO(1-0) observations made with the Nobeyama Millimeter Array at a resolution of $2\farcs5$ (50 pc) show a velocity dispersion of $89$ km s$^{-1}$ \citep{CM2016}. \citet{T2018} measured the velocity dispersion of individual clumps in the disk of M82 to be $\sim100$ km s$^{-1}$ using 0\farcs05-resolution CO observations with ALMA. They also estimated a velocity dispersion of $74\pm1$ km s$^{-1}$ for the CO-emitting gas within a disk of $\sim1$ kpc radius. Understanding that the velocity dispersion of the disk shows a very complex structure, we find that the most complementary observations for our analysis come from \citet{Leroy2015}. They show that the molecular gas, $^{12}$CO(2-1), traces the high-density regions of M82, which is spatially coincident with the dust emission from our FIR observations. We use their $^{12}$CO(2-1) emission line within our mask, which we smoothed using a Gaussian profile with a FWHM equal to the beam size of the HAWC+ observations and projected to the HAWC+ observations. We estimate a median velocity dispersion of $\sigma_{v,^{12}\rm CO(1-2)} = 66.0\pm6.6$ km s$^{-1}$ (Fig. \ref{fig:fig2}).

The final variable required to estimate $B_{\rm DCF}$ is $\sigma_{\phi}$, which corresponds to the standard deviation of the polarization angle distribution. Here, we use the polarization angle of the inferred B-field orientation from the 53 \um\ HAWC+ observations \citep{Jones2019} within our mask to estimate a median value of $-15.3^{\circ}$ with an angular dispersion (std. deviation) of $\sigma_{\phi} = 17.1^{\circ}$. Within the mask, $17$\% of polarization measurements have a $P/\sigma_{P}$ between 2 and 3, with a median $P/\sigma_{P}= 8$. Thus the angular dispersion is larger than the uncertainty related to the polarization measurement. Therefore, using $\sigma_{\phi}$ = 0.29 rad (17.1$^{\circ}$), we find $B_{\rm DCF} = 0.54 \pm0.17$ mG.


\subsection{The Angular Dispersion Function}\label{subsec:Badf}

In the starburst region of M82, the  dominant driver of the turbulence is the supernova explosions. The observed patterns of the B-field orientation are the results of two contributions: 1) one from the morphology of the large-scale regular magnetic field, which is larger than the turbulent scale ($\delta$) driven by supernova explosions at scales of $50-100$ pc in nearby galaxies \citep[i.e.][]{Ruzmaikin1988,Brandenburg2005,ES2004,Haverkorn2008}; and 2) another contribution from the small-scale ({\it i.e.,} turbulent or random) magnetic field, which relies on turbulent gas motions at scales compared to or smaller than $\delta$.  Because the DCF method relies on the speed of an Alfv\'en wave to measure the magnetic field strength, only the perturbed ({\it i.e.,} turbulent or random) component  provides the correct value of $B$. Therefore, it is important to extract the turbulent component from the measured dispersion.

\citet{H2009} and \citet{Houde2009, Houde2011} have been able to separate the regular and turbulent components using a careful analysis of the dispersion of polarization angles obtained from dust continuum polarization observations. This technique has been applied to FIR-sub-mm polarimetric observations of Galactic sources \citep[i.e.][Li et al. in prep.]{Chapman2011,Crutcher2012,Girart2013,Pattle2017,Soam2019,Chuss2019,Redaelli2019,Wang2020,Guerra2021,Michail2021}, as well as sub-mm polarimetric observations of external galaxies, like M51 \citep{Houde2013}. Specifically, an isotropic two-point structure function ({\it i.e.,} dispersion function) is computed to describe the dispersion as a function of angular scale. Then, the dispersion function separates the large-scale field from that of the turbulence \citep[][eq. 13]{Houde2016},


\begin{widetext}
\begin{equation}
1 - \langle \cos[\Delta\phi(l)] \rangle = \frac{1}{1+\mathcal{N} \left[ \frac{\langle B_{t}^{2} \rangle}{\langle B_{0}^{2} \rangle} \right]^{-1}}  \left\{ 1 - \exp \left( -\frac{l^2}{2(\delta^2+2W^2)} \right) \right\} + a_{2}l^2
\label{eq:eqadf}
\end{equation}
\end{widetext}

\noindent
where the first term on the right accounts for the small-scale turbulent contribution to the dispersion, and the second term accounts for the large-scale regular (ordered) field contribution. $l$ is the distance between pairs of measurements with angle difference $\Delta\phi$. $W$ is the standard deviation of the beam size assumed to be a Gaussian function $W = $ FWHM$/2.355 = 2\farcs05$, where FWHM = $4\farcs85$ for our 53 \um\ HAWC+ observations \citep{Harper2018}. $\langle B_{t}^{2} \rangle / \langle B_{0}^{2} \rangle$ is the turbulent-to-large-scale energy ratio, $\delta$ is the correlation length for the turbulent field, $a_{2}$ is the large-scale coefficient, and $\mathcal{N}$ is the number of independent turbulent cells in the column of dust probed observationally given by 

\begin{equation}
\mathcal{N} = \frac{(\delta^2 + 2W^{2})\Delta'}{\sqrt{2\pi}\delta^3}
\label{eq:eqN}
\end{equation}

\noindent
\citep[][eq. 14]{Houde2016}, where $\Delta'$ is the effective thickness of the starburst region, which was already estimated in Section \ref{subsec:Bclassic}. 

\begin{figure*}[ht!]
\includegraphics[angle=0,scale=0.6]{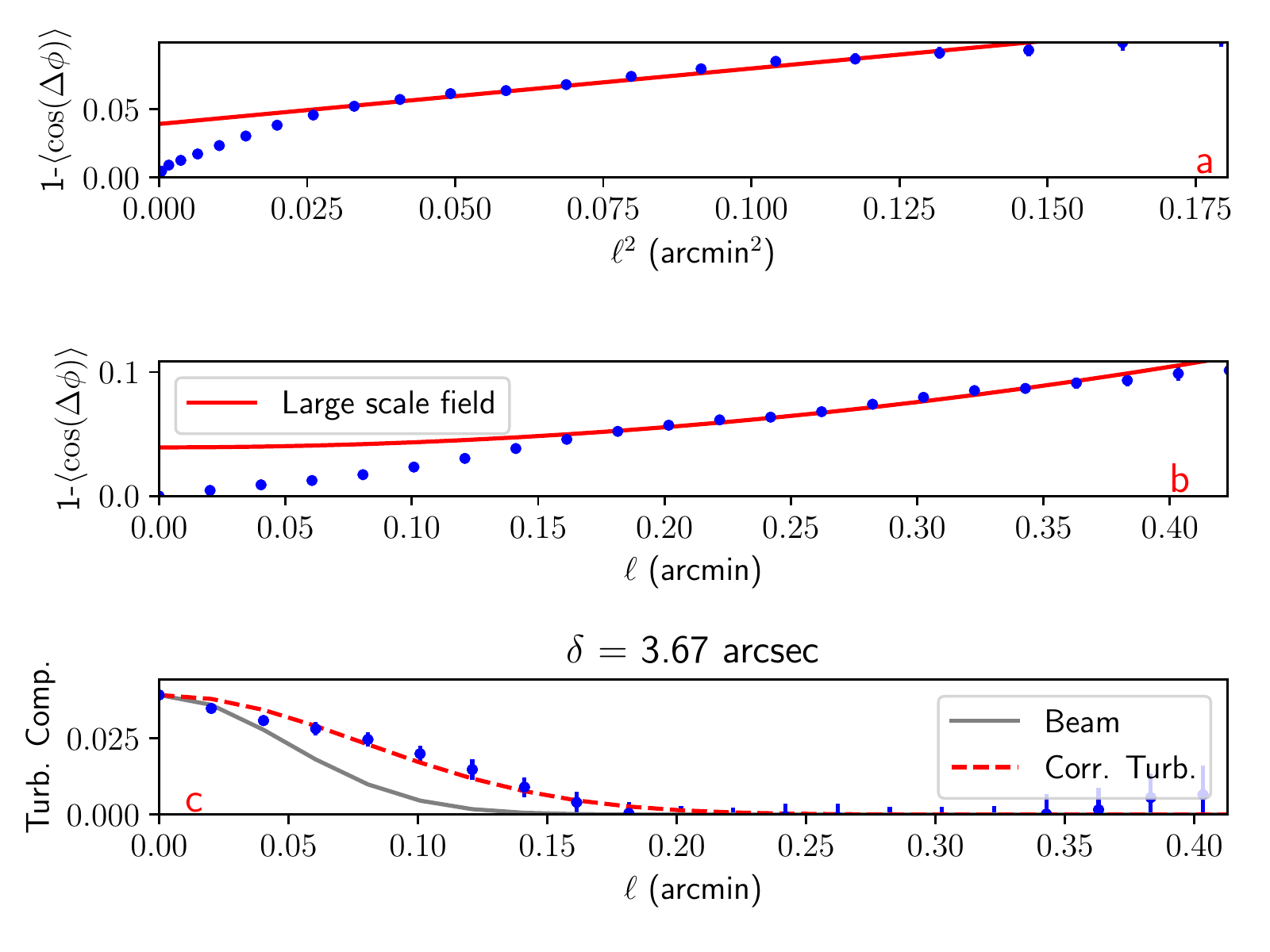}
\includegraphics[angle=0,scale=0.4]{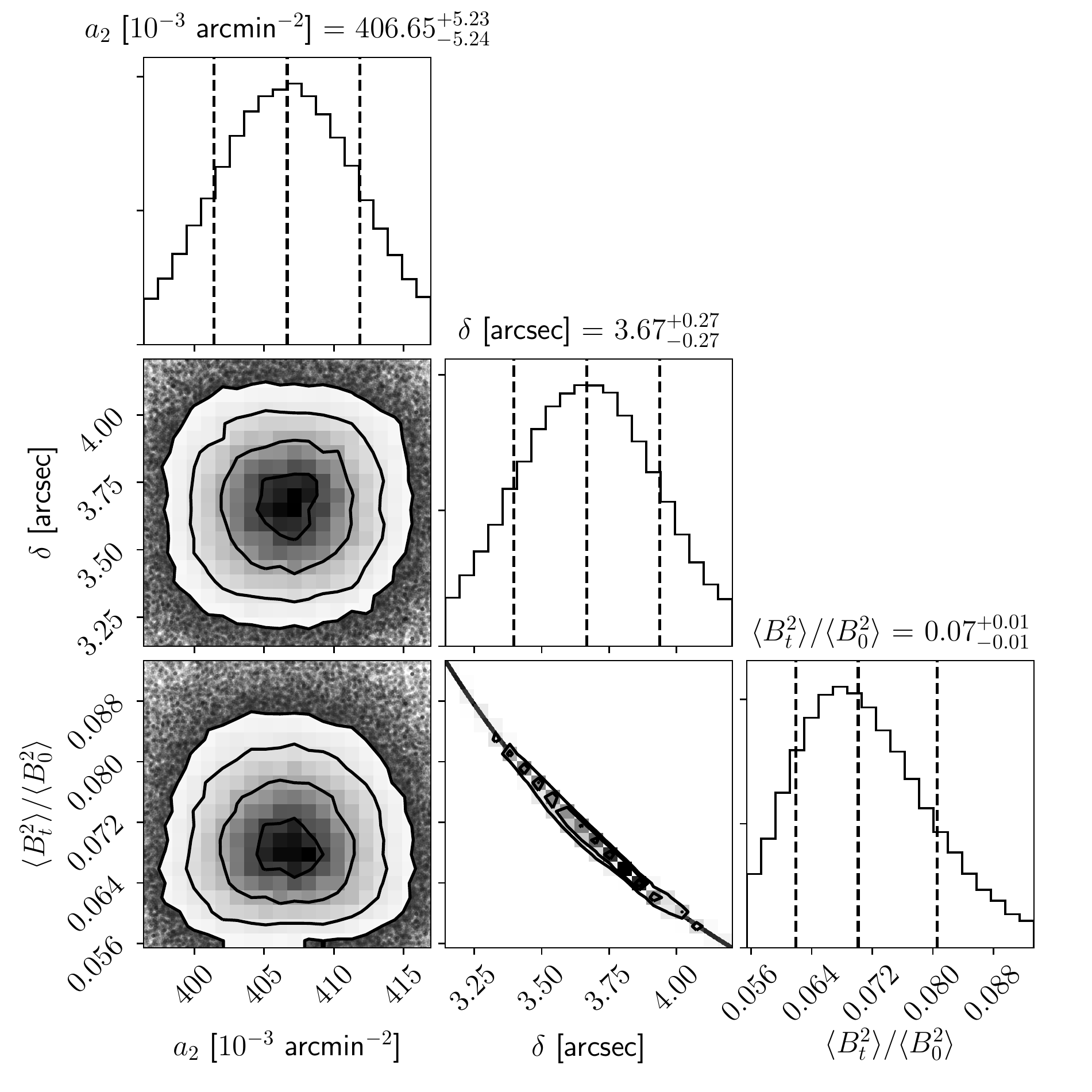}
\caption{\textit{Left:} Dispersion function (a,b) within the starburst mask of M82 at 53 \um. Data points (blue circles) and fits (red solid line) of the large-scale field are shown. (c) The best fit (red dashed-line) of the turbulent component and the beam (grey solid line) of the observations are shown. The fits represent the best inferred results using Eq. \ref{eq:eqadf}.
\textit{Right:}  Posterior distributions of the large-scale field ($a_{2}$), turbulent correlation length ($\delta$), and turbulent-to-large-scale field ratio ($\langle B_{t}^{2} \rangle / \langle B_{0}^{2} \rangle$).
\label{fig:fig3}}
\epsscale{2.}
\end{figure*}

We evaluate the left-hand side of Eq. \ref{eq:eqadf} within the starburst mask using the 53 \um~polarization data of M82 observed with HAWC+ \citep{Jones2019}. Using $\Delta' = 0\farcm1613$ ($193.6$ pc), this leaves three free parameters to be determined: $\ratio$, $\delta$, and $a_{2}$. We use a Markov Chain Monte Carlo (MCMC) solver for fitting the non-linear model of Eq. \ref{eq:eqadf} to the dispersion functions as implemented by \citet{Chuss2019}. This fitting routine infer the optimal model parameters and its associated uncertainties from the posterior distributions of the MCMC chains. Figure \ref{fig:fig3}-left shows the separation of the two-components in the measured dispersion function $1 - \langle \cos[\Delta\phi(l)] \rangle$. Panels a and b show in blue circles the measured values of $1 - \langle \cos[\Delta\phi(l)] \rangle$ and the best-fit (red solid line) for the large-scale component (term $\propto\ell^{2}$ in Eq.\ref{eq:eqadf}), which is valid only for $\ell \gtrsim 0.18$ arcmin. For $\ell \le 0.18$ arcmin, the turbulent component dominates which is then analyzed in panel c. In panel c, blue circles correspond to the autocorrelated turbulent component ($1 - \langle \cos[\Delta\phi(l)] \rangle-a_{2}\ell^{2}$), which is valid only for the smaller scales and is fit (dashed red line) using the first term in Eq.\ref{eq:eqadf}. The turbulent component is compared with autocorrelated beam profile (solid grey line). Correctly accounting for the gas turbulence in the polarization angle dispersion depends on the autocorrelated turbulent component having a wider shape than the observations' autocorrelated beam. This is evident in Figure \ref{fig:fig3}c. Figure \ref{fig:fig3}-right shows the posterior distributions obtained with the MCMC solver, which statistics provide the best-fit model parameter values for the dispersion function with $a_{2} = 406.65^{+5.23}_{-5.24} \times 10^{-3}$ arcmin$^{-2}$.

We infer a coherent length of the turbulent magnetic field to be $\delta = 3\farcs67^{+0.27}_{-0.27}$ ($73.6^{+5.6}_{-5.6}$ pc). The coherent length is larger than the beam size of our observations (Figure \ref{fig:fig3}c), $\sqrt{2}W$ = 2\farcs90, which allow the characterization of the turbulent magnetic field \citep{Houde2011}. It is worth noticing that our coherent length of the turbulent magnetic field is in agreement with the typical scale length of turbulent fields in galaxies of order $50-100$ pc, which is driven by supernova explosions \citep[i.e.][]{Haverkorn2008}. \citet{Adebahr2017} estimated a coherent length of the magnetic field of $\sim50$ pc using radio polarimetric observations in the central $\sim 1\times0.7$ kpc$^{2}$ of M82, in agreement with our results. Using isotropic Kolmogorov turbulence ($P(k) \propto k^{5/3}$), we estimate a coherent length \citep{LC2018} of $L_{c} = \delta/5 = 14.7$ pc, where $\delta$ is considered to be the maximum coherent length, $L_{max}$, within the starburst region. We conclude that our observations are able to resolve the turbulent magnetic field in the central $873 \times 510$ pc$^{2}$ starburst of M82 and therefore, further detailed analysis of the magnetic field strengths can be performed.

We infer a turbulent-to-large-scale energy of $\ratio = 0.07^{+0.01}_{-0.01}$. This result implies that the central starburst region of M82 is dominated by a large-scale ordered magnetic field since the turbulent magnetic energy is small ($\approx$7\%) in comparison. We quantify the effect of the galactic outflow in Section \ref{subsec:Boutflow}.

We can now estimate the POS magnetic field strength using the angular dispersion function (ADF) as

\begin{equation}\label{eq:BADF}
B_{\rm ADF}  = \sqrt{4\pi\rho} \sigma_{v} \left[ \frac{\langle B^{2}_{t} \rangle}{\langle B_{0}^{2} \rangle} \right]^{-1/2}
\end{equation}

\noindent
\citep{Houde2009}. The values of $\rho$, the mass density, and $\sigma_{v}$, the velocity dispersion, are those previously estimated in Section \ref{subsec:Bclassic} and Fig. \ref{fig:fig2}. Using these parameters, we estimate the magnetic field strength within the starburst mask of M82 to be $B_{\rm ADF} = 1.04\pm0.17$ mG.


\subsection{The Effect of Galactic Outflows}\label{subsec:Boutflow}

In this section, we investigate how both values of the magnetic field, $B_{\rm DCF}$ and $B_{\rm ADF}$, are affected by the M82 outflow. Assuming there is a steady, large-scale velocity field $\vec{U}_{0} = U_{0}\hat{z}$ in the same direction as the magnetic field $\vec{B}$, the wave equation for this case -- after several steps -- reduces to \citep[Eq. 10 by ][]{Nakariakov1998}:

\begin{equation}
    \left[\frac{\partial}{\partial t}+(U_{0} - V_{A})\frac{\partial}{\partial z} \right]\left[\frac{\partial}{\partial t}+(U_{0} + V_{A})\frac{\partial}{\partial z} \right]V = 0.
\end{equation}

This means there are two standing waves with velocities $V^{-} = U_{0} - V_{A}$ and $V^{+} = U_{0} + V_{A}$. However, these two velocities are associated with the same observed spatial disturbance ({\it i.e}, polarization-angle dispersion). Replacing this modified velocity into Eq. \ref{eq:V2},

\begin{equation}
    (V^{\pm})^{2} = (U_{0} \pm V_{A})^{2} = \frac{\sigma_{v}^{2}}{\sigma_{\phi}^{2}},
\end{equation}

\noindent
results in

\begin{equation}
      \pm V_{A} = \frac{\sigma_{v}}{\sigma_{\phi}} - U_{0},
\end{equation}

\noindent
Using the definition of Alfv\'en wave (Eq. \ref{eq:Va}),

\begin{equation}
    \pm \frac{B}{\sqrt{4\pi\rho}} = \frac{\sigma_{v}}{\sigma_{\phi}} - U_{0}, \label{eq:B_plus_minus}
\end{equation}

\noindent
which indicates that two Alfv\'en waves with the same speed travel in opposite directions, due to the same magnetic field strength and the same non-zero, steady-state large scale velocity. Therefore, we can take the absolute value of Eq. \ref{eq:B_plus_minus} and re-arrange the terms,

\begin{equation}
    B = \sqrt{4\pi\rho}\left|\frac{\sigma_{v}}{\sigma_{\phi}} - U_{0}\right|,
\end{equation}

\noindent
where $B$ is the absolute value of the field strength. Finally, using the definition of $B_{\rm DCF}$ (including the adjustment factor $\xi$), we define the modified $B^{\prime}_{\rm DCF}$ as

\begin{equation}\label{eq:Bdcfp}
    B^{\prime}_{\rm DCF} = B_{\rm DCF}\left|1 - \sigma_{\phi}\frac{U_{0}}{\sigma_{v}}\right|.
\end{equation}

\noindent
Equation \ref{eq:Bdcfp} reduces to the well-known DCF approximation (Eq. \ref{eq:DCF}) when $U_{0}=0$. Assuming, of course, that $\sigma_{\phi}$ is non-zero, the modification to the DCF value is proportional to $U_{0}/\sigma_{v}$ --- the ratio of large-scale velocity to the velocity dispersion. 

From Eq. \ref{eq:Bdcfp} we now have two possible regimes:

\begin{itemize}
    \item $B^{\prime}_{\rm DCF} < B_{\rm DCF}$, which means $\sigma_{\phi}\frac{U_{0}}{\sigma_{v}} < 2$ and $\neq$ 1.
    
    \item $B^{\prime}_{\rm DCF} > B_{\rm DCF}$, which means $\sigma_{\phi}\frac{U_{0}}{\sigma_{v}} > 2.$
\end{itemize}

\noindent
Therefore the correction to the DCF value (Eq. \ref{eq:Bdcfp}) increases the strength of the magnetic field when the ratio of dispersion-to-large-scale  velocities is lower than half the measured angle dispersion ($0.5\sigma_{\phi}>\sigma_{v}/U_{0}$) --- the large-scale velocity field dominates. On the other hand, the correction lowers the magnetic field strength when the turbulence dominates the velocity field ($0.5\sigma_{\phi}<\sigma_{v}/U_{0}$).

Using the median velocity of the $^{12}$CO(2-1) molecular outflow within the starburst mask of $U_{0} = 396\pm87$ km s$^{-1}$ \citep{Leroy2015}, the $\sigma_{\phi}=$ 0.29 rad., and $\sigma_{v}=$66.0$\pm6.6$ km s$^{-1}$ from Section \ref{subsec:Bclassic}, we estimate the modifying factor to be:

\begin{equation}
    \sigma_{\phi}\frac{U_{0}}{\sigma_{v}} = 1.74\pm0.42
\end{equation}

\noindent
since this factor is $<$2, we conclude that the large-scale velocity field from the galactic outflow dominates over the turbulent magnetic field within the starburst mask. This result is in agreement with that the large-scale regular magnetic field dominates over the turbulent magnetic energy within the starburst region. These results imply that the $B_{\rm DCF}$ and $B_{\rm ADF}$ are overestimated. We can now calculate the corrected strength of the magnetic field within the starburst mask for both the classical DCF method, $B^{\prime}_{\rm DCF} = 0.40\pm0.26$ mG, and the angular dispersion analysis method, $B^{\prime}_{\rm ADF} =  0.77\pm0.17$ mG.

Both estimates of the magnetic field, $B^{\prime}_{\rm DCF}$ and $B^{\prime}_{\rm ADF}$, agree within the uncertainties when a galactic outflow component is taken into account. The separation of turbulent-to-large-scale fields, as described in \cite{Houde2009}, does take into account some influence of the large-scale velocities present in the starburst mask. However, the ADF method still requires a correction of $\sim25$\%. As the corrected ADF method is now dominated by the turbulent field, we use $B^{\prime}_{\rm ADF}$ hereafter.


\section{The two-dimensional map of the magnetic field strength}\label{sec:2dB}

In Section \ref{sec:Bfield}, we estimated the average turbulent magnetic field strength in the starburst region of M82 using the mean of the mass density, velocity dispersion, and angular dispersion. Here, we estimate the pixel-by-pixel turbulent magnetic field strength using the full maps of the mass density and velocity dispersion. A similar approach has successfully been applied to the Orion nebula (OMC-1) by \citet{Guerra2021}. We here present a novel approach using the energy budget to estimate the two-dimensional turbulent magnetic field strength of M82.

\citet{Contursi2013} suggested that the cold clouds from the disk are entrained into the outflows by the galactic wind, where small, dense clouds may remain intact during the outflow process. \citet{Jones2019} proposed a scenario where the magnetic field is entrained within the gas and dust, and dragged by the outflow away from the galactic disk together with the galactic wind.  This entrainment is generally quantified by the plasma $\beta$ parameter, defined as the ratio of thermal gas pressure, $U_{G}$, to magnetic pressure, $U_{B}$. $\beta$ traditionally determines whether an environment is dominated by thermal or magnetic forces:

\begin{equation}\label{eq:plasmab}
\beta = \frac{U_{G}}{U_{B}} = \frac{n_{H} k_{B} T_{g}}{B^{2}/8\pi}
\end{equation}
\noindent
where n$_{H}$ is the gas density, $k_{B}$ is the Boltzmann constant, and T$_{g}$ is the gas temperature.

Turbulent energy densities are generally larger than thermal energy densities in galaxies \citep{Beck2005}. Thus, a $\beta-$like parameter that takes both the turbulent kinetic and hydrostatic energies of the outflow of M82 into account is required. We here define a $\beta'$ parameter:

\begin{equation}\label{eq:bprime}
    \beta' = \frac{U_{H}+U_{K}}{U_{B}}
\end{equation}
\noindent
where $U_{H}$, $U_{K}$, and $U_{B}$ are the hydrostatic, turbulent kinetic, and turbulent magnetic energies, respectively.

Let the hydrostatic energy, $U_{H}$, be 

\begin{equation}\label{eq:UH}
    U_{H} = \pi G \Sigma_{g}^{2} = \pi G (N_{H} m_{H} \mu )^{2}
\end{equation}
\noindent
where $G$ is the gravitational constant, $\Sigma_{g}$ is the gas density, $N_{H}$ is the gas column density, $m_{H}$ is the hydrogen mass, and $\mu = 2.8$ is the mean molecular mass per H molecule.  

Let the turbulent kinetic energy, $U_{K}$, and magnetic energy, $U_{B}$, be

\begin{equation}\label{eq:UK}
    U_{K} = \frac{1}{2}\rho\sigma_{v}^{2} 
\end{equation}
\begin{equation}\label{eq:UB}
    U_{B} = \frac{B^{2}}{8\pi}
\end{equation}
\noindent
where $\sigma_{v}$ is the three-dimensional dispersion velocity defined as $\sqrt{3}\sigma_{v,^{12}CO(2-1)}$. 

Using $\beta'$ (Eq. \ref{eq:bprime}), the two-dimensional magnetic field strength map can be estimated such as

\begin{eqnarray}\label{eq:B}
    B &=& \left[8\pi\left(\frac{U_{K} + U_{H}}{\beta'}\right)\right]^{1/2} \nonumber \\
      &=& \left[ \frac{8\pi}{\beta'}(\pi G N_{H} m_{H} \mu + \frac{1}{2}\rho\sigma_{V}^{2})  \right]^{1/2}
\end{eqnarray}

\noindent
where we impose the condition that $\beta'$ is equal to the mean value of the energies within the starburst mask.

\begin{figure*}[ht!]
\includegraphics[angle=0,scale=0.42]{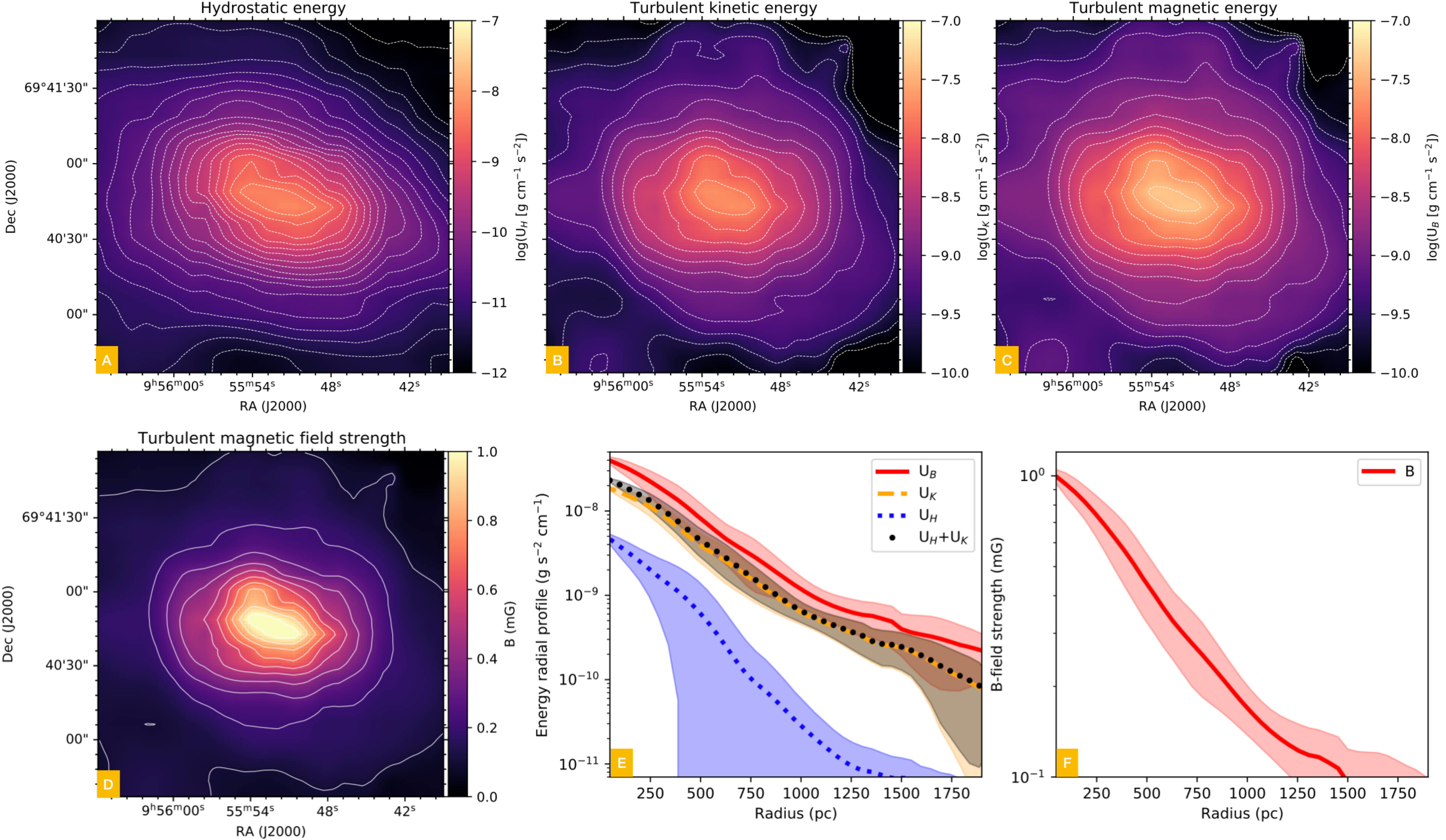}
\caption{Hydrostatic (A), turbulent kinetic (B), and magnetic (C) energy maps within a $140\arcsec \times 140\arcsec$ ($2.8 \times 2.8$ kpc$^{2}$) region. Contours start at $-12$ and increases in steps of $0.8$ in logarithmic scale. (D) Turbulent magnetic field strength map. Contours start at $0.1$ mG and increase in steps of $0.1$ mG. Radial profiles of the energies (E) and turbulent magnetic field strength (F) showing the median (lines) and 1$\sigma$ uncertainties (shaded region).
\label{fig:fig4}}
\epsscale{2.}
\end{figure*}

We use the mean values within the starburst mask from Section \ref{sec:Bfield}, and estimate a $\beta' = 0.56\pm0.23$. Imposing this condition satisfies that the estimated mean turbulent magnetic field strength within the starburst mask is $\langle B \rangle = B_{\rm ADF}^{\prime} = 0.77$ mG. In combination with the maps of $N_{H+H_{2}}$, $\sigma_{v,^{12}\rm CO(2-1)}$ from Section \ref{sec:Bfield}, we compute the two-dimensional maps of the energies and turbulent magnetic field strength of M82. Figure \ref{fig:fig4}-A-D shows the resulting two-dimensional maps of the hydrostatic, turbulent kinetic, and turbulent magnetic energies as well as the turbulent magnetic field strength, each $140\arcsec \times 140\arcsec$ ($2.8\times2.8$ kpc$^{2}$) in size. Figure \ref{fig:fig4}-E-F shows the radial profiles of the energies and turbulent magnetic field strength. We estimate the median and standard deviation of the median (1$\sigma$) for annulus of width equal to the Nyqvist sampling ($2\farcs43$, $48.5$ pc) of the HAWC+ observations.

We find that the turbulent kinetic and turbulent magnetic energies are in close, within 1$\sigma$, equipartition across the galactic outflow at radius $\sim 300-1500$ pc. At radius $\le 300$ pc, the turbulent magnetic energy appears to be higher than the turbulent kinetic energy. We note that this result within the starburst mask may imply that  1) the observed turbulent magnetic energy may be larger than the combined turbulent kinetic and hydrostatic energy in the compact star-forming regions, or 2) there may be several magnetic field components that may be overestimating the median turbulent magnetic field strength. A decomposition of the magnetic field strengths into small-scale turbulent fields (i.e. anisotropic random fields) due to star-forming regions or shocks in the superwind, and large-scale fields due to galactic superwind are required to quantify the contribution of these several components to the observed magnetic field strength within the starburst mask (see Section \ref{sec:PFvsB}). At radius $>1500$ pc, the turbulent kinetic energy shows a steeper drop than the turbulent magnetic energy, however both energies still remain close to equipartition, within $1\sigma$ uncertainties.

\citet{Thompson2006} estimated the magnetic field strengths of starburst galaxies using a hydrostatic approach. These authors argued that the starbursts' magnetic field strengths are much larger than the magnetic field strengths inferred using the observed radio fluxes and assuming equipartition between cosmic rays and magnetic energy densities. These results are derived if the cosmic ray electron cooling timescale is shorter than the escape timescale from the galactic disk. This work assumed that only the hydrostatic pressure takes place in the energy balance with the magnetic energy, which results in the scaling  $B \propto \Sigma^{a}_{g}$, where $\Sigma_{g}$ is the surface brightness of the gas, and $a$ is the power-law index. $a =0.7$ if the magnetic energy is in equipartition with the pressure from star formation in the ISM, and $a =0.4$ for equipartition between cosmic rays and magnetic energy densities. Note that our magnetic field strength (Eq. \ref{eq:B}) depends on the turbulent kinetic and hydrostatic energies, and it does not assume equipartition ($\beta^{\prime}$ is included).

\section{Potential Field Extrapolation}\label{sec:PF}

We are interested in quantifying the influence of the galactic outflow in the magnetic field towards the intergalactic medium. This study requires the knowledge of the magnetic and kinetic energies at distances of several kiloparsecs from the galaxy plane. Although the kinetic energy can be estimated up to tens of kpc, there are no empirical measurements to compute the magnetic energies at these distances in M82. Our magnetic field results of M82 obtained in Section \ref{sec:2dB} provide an observational boundary condition required to perform a potential field extrapolation and estimate the field strength and morphology at several kpc above and below the galactic plane.

We employ a standard, well-tested technique that has been used in solar physics for decades to determine the magnetic field in the solar corona, a region where the magnetic field (with rare exceptions) cannot be measured directly. In the solar physics case, the LOS magnetic field is measured in the photosphere using the Zeeman Effect. This observation provides the first required boundary condition; the second condition assumes the field reduces to zero at infinity \citep[see {\it e.g.},][]{Sakurai1982}. For a comprehensive review of magnetic field extrapolation techniques in solar physics please see \citet{Wiegelmann2012}. Potential field extrapolation has also been employed to study the magnetic field in slow-rotating and cool stars \citep{Donati2001,Hussain2001} such as $\tau$ Sco \citep{Donati2006}. Here, we have modified the solar physics method to work with POS polarization data from HAWC$+$.

\subsection{Solving Laplace's Equation}\label{subsection:PF_def}

The simplest of these extrapolation approximations assumes that the electrical currents are negligible. In this case, the magnetic field, $B = -\nabla \phi$, has a scalar potential, $\phi (x,y,z)$, that satisfies the Laplace equation \citep[see {\it e.g.}][]{Neukirch2005}

\begin{equation}
{\nabla}^2 \Phi = 0.\label{laplace}
\end{equation}

\noindent
Here, the plane $x-y$ is parallel to the galaxy's disk and our extrapolation will be limited to the $x-z$ plane, above and below the galactic plane. This two-dimensional Cartesian geometry requires one boundary condition from the magnetic field map (Section \ref{sec:2dB}) along the plane of the galaxy's disk at $xz=(x,0)$ 

\begin{equation}
B_z(x,0) = F(x) \label{bc}
\end{equation} 

\noindent
and a second boundary condition at infinity, $B_z  \rightarrow 0$ as $|z| \rightarrow \infty$. Using a separation of the variables

 \begin{equation}
 \Phi (x,z) = X(x) Z(z).
 \label{sep}
 \end{equation}
 
\noindent
Substituting Eq. \ref{sep} into Eq. \ref{laplace} gives 

 \begin{equation}
X^{\prime \prime} Z + X Z^{\prime \prime} = 0 .
 \end{equation}
 
\noindent
A particular solution with wavenumber $k$ has $X^{\prime \prime} = -k^2 X$ and $Z^{\prime \prime} = k^2 Z$, so 

\begin{eqnarray}
X(x) &=& A \cos(kx) + B \sin(kx),\\
Z(z) &=& C e^{-k|z|} + D e^{k|z|} ,
\end{eqnarray}

\noindent
where $A$, $B$, $C$, and $D$ are constants. We now have

\begin{equation}
B_z = \frac{\partial \Phi}{\partial z} = X(x) (-C k e^{-k|z|} + D k e^{k|z|}),
\end{equation}

\noindent
As $B_z  \rightarrow 0$ when $z \rightarrow \infty$, the constant $D = 0$. We can also set $C=1$ as the constants can be set by the values of $A$ and $B$. Then $Z(z) =e^{-k|z|}$,  $ X(x)= -F(x)/k$, and

\begin{equation}
\Phi_k = -\frac{1}{k}F(x)e^{-k|z|}.
\end{equation}

\noindent
We will assume that the source of $\vec B$ is finite, so that $B_z(x,0) = F(x)=0$ for $|x|> \ell/2$ for some length $\ell$. 
We will also assume that the net flux into the upper half plane is zero, i.e.

\begin{equation}
\int_{-\ell/2}^{\ell/2} F(x) dx = 0.
\end{equation} 

\noindent
Expansion into a Fourier series gives 

\begin{equation}
F(x) = \sum_{n=1}^{n=\infty}(a_n \cos k_nx + b_n \sin k_n x) , \qquad k_n = \frac{2\pi n}{\ell}.
\end{equation}

\noindent
We now have

\begin{eqnarray}
B_x(x,z) &=& \sum_{n=1}^{n=\infty}(a_n \sin k_nx - b_n \cos k_n x)e^{-k_n|z|}
\label{eq:pot_Bx}
\end{eqnarray}
\begin{eqnarray}
B_z(x,z) &=& \sum_{n=1}^{n=\infty}(a_n \cos k_nx + b_n \sin k_n x)e^{-k_n|z|}.
\label{eq:pot_Bz}
\end{eqnarray}

\noindent
The coefficients are given by

\begin{eqnarray}
a_n &=& \frac{2}{\ell}\int_{-\ell/2}^{\ell/2} F(x) \cos(k_n x) dx,
\label{eq:coeff_a}
\end{eqnarray}
\begin{eqnarray}
b_n &=& \frac{2}{\ell}\int_{-\ell/2}^{\ell/2} F(x) \sin(k_n x) dx.
\label{eq:coeff_b}
\end{eqnarray}

Finally, the potential field orientations in the POS are estimated as

\begin{equation}
    \phi = {\rm arctan}\left(\frac{B_{z}}{B_{x}}\right),
    \label{eq:pot_ang}
\end{equation}
\noindent
where the magnetic field components require to be projected on the observer's view given the inclination and tilt angle of the galaxy on the POS.

\subsection{Potential Field Solutions}\label{subsection:PF_res}

\begin{figure}[ht!]
\centering
\includegraphics[angle=0,scale=0.45]{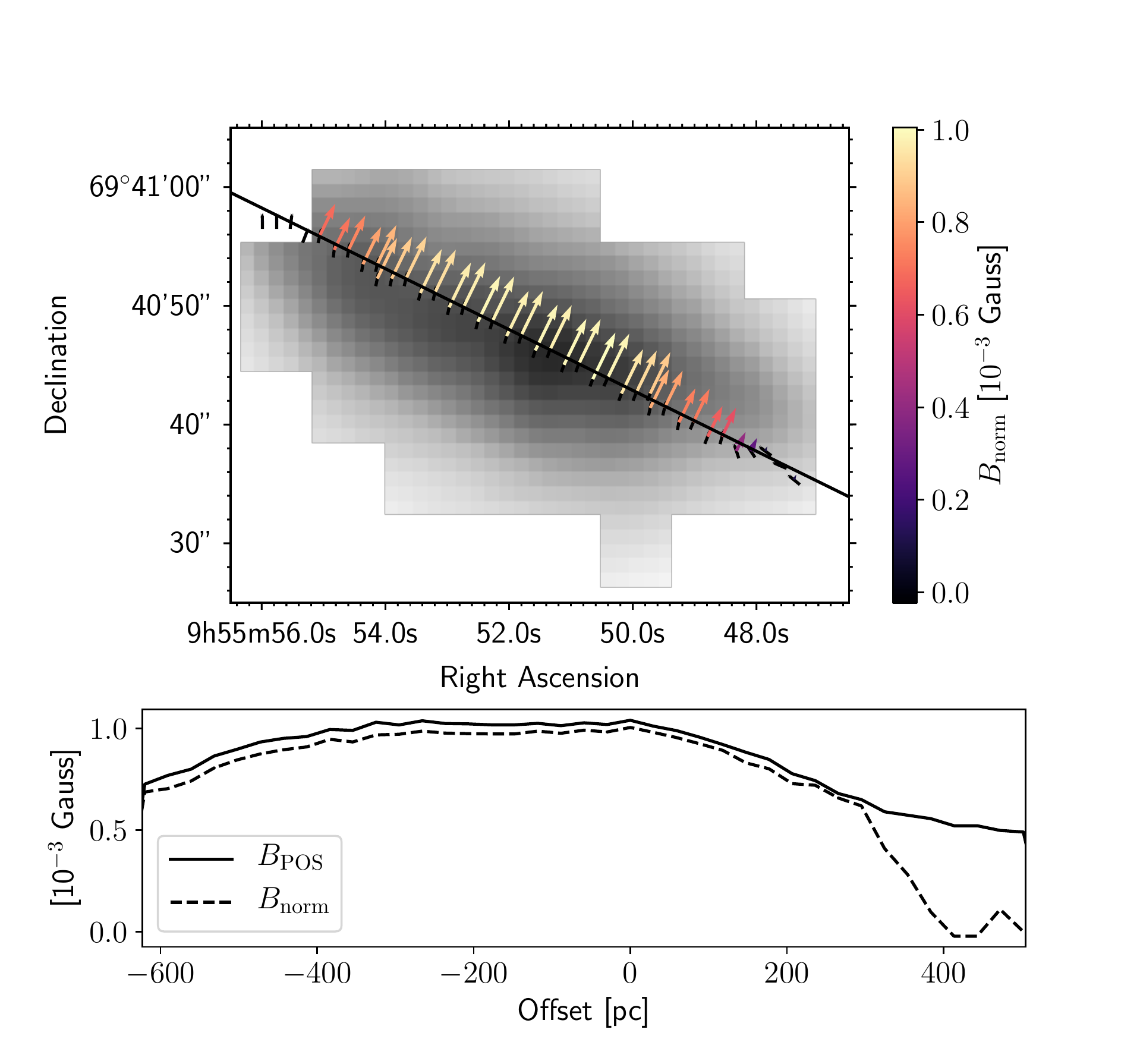}
\caption{ {\it Top:} Boundary values $B_{\rm Norm}$ for the potential field extrapolation in M82. $B_{\rm Norm}$ vectors length and color correspond to the $B_{\rm POS}$ strength. The solid black line represents the plane of the galaxy defined at tilt angle 64$^{\circ}$ \citep{Mayya2005}. HAWC+ rotated polarization orientations (black) and arrows corresponding to the normal-to-plane component of the magnetic field (red) are shown along the galaxy's plane. Background corresponds to the total intensity at 53 \um. {\it Bottom:} $B_{\rm POS}$(solid) and $B_{\rm Norm}$(dashed) profile as a function of the offset position from the maximum $I$ intensity along the galaxy's plane.}
\label{fig:fig5}
\epsscale{2.}
\end{figure}

The boundary condition, $F(x)$, corresponds to the normal-to-plane component of the magnetic field ($B_{\rm Norm}$), so the observed POS magnetic field ($\vec{B}_{\rm POS}$) needs to be reprojected. We define the galaxy's plane by its tilt angle of $\theta=64^{\rm o}$ measured as positive in the East from North direction \citep{Mayya2005}; this is designated by the black solid line in Figure \ref{fig:fig5}. $\vec{B}_{\rm POS}$ corresponds to the measurements of the magnetic field strength (Figure \ref{fig:fig4}-D) and orientation (pseudo-vectors in Figure \ref{fig:fig5}) along this line. We use Euler rotations around the x-axis, $R_{x}[i]$, and the z-axis, $R_{z}[\theta]$, to compute $\vec{B}_{\rm Norm} = R_{x}[i]R_{z}[\theta]\vec{B}_{\rm POS}$ using an inclination angle of $i=76^{\circ}$ \citep{Mayya2005} with respect to the LOS. $\vec{B}_{\rm Norm}$ is displayed in Figure \ref{fig:fig5} as vectors along the galaxy's plane with lengths and color corresponding to their reprojected strength. In Figure \ref{fig:fig5} both $B_{\rm POS}$ and $B_{\rm Norm}$ strengths are seen as a function of the offset radius from the maximum of total intensity along the galaxy's plane.

It is important to recall here that magnetic field orientations determined by FIR polarimetric data suffer from the 180$^{\circ}$ ambiguity \citep{H2000} -- that is all vectors in Figure \ref{fig:fig5}-{\it Top} have the same likelihood to be pointing into the galaxy's plane. This angular ambiguity does not affect the shape or strength of the magnetic field lines, just the direction. Thus, we choose to calculate the potential field solution with this choice of direction for $B_{\rm Norm}$ and only work with the orientation of the potential solutions.

\begin{figure*}[ht!]
\centering
\includegraphics[angle=0,scale=0.65]{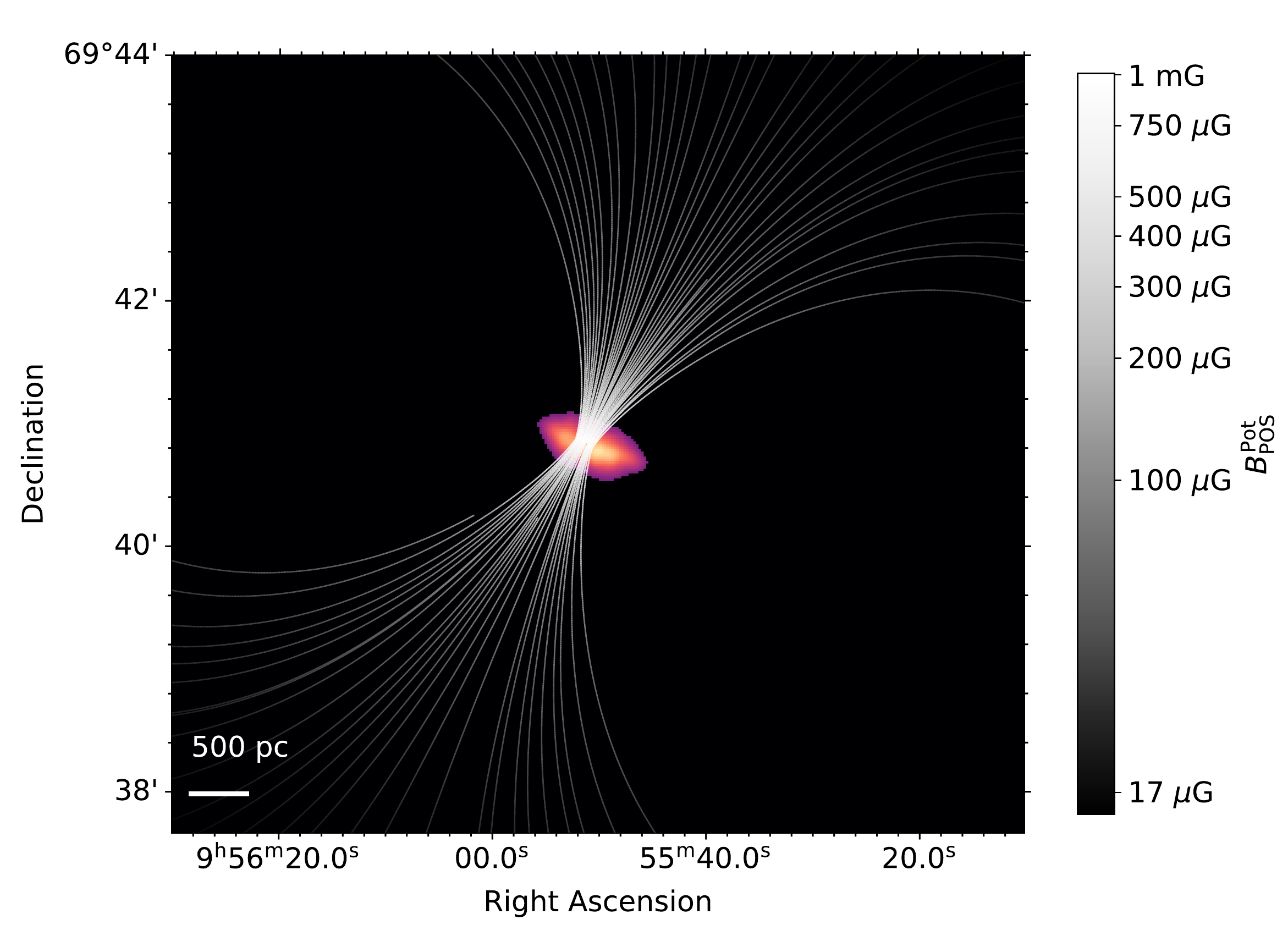}
\caption{Potential magnetic field lines of M82 inferred using the $53$ \um~polarimetric observations with HAWC+/SOFIA. The potential magnetic field is calculated by extrapolation of the magnetic field at the galaxy's plane. Magnetic field lines are visualized in a field of view $\sim$7.2 kpc$^{2}$ centered at M82. The potential field strength is larger in the bulk of the galaxy in good agreement with the 2D map of M82 in Figure \ref{fig:fig4}.
\label{fig:fig6}}
\epsscale{2.}
\end{figure*}

\begin{figure}[ht!]
\centering
\includegraphics[angle=0,scale=0.35]{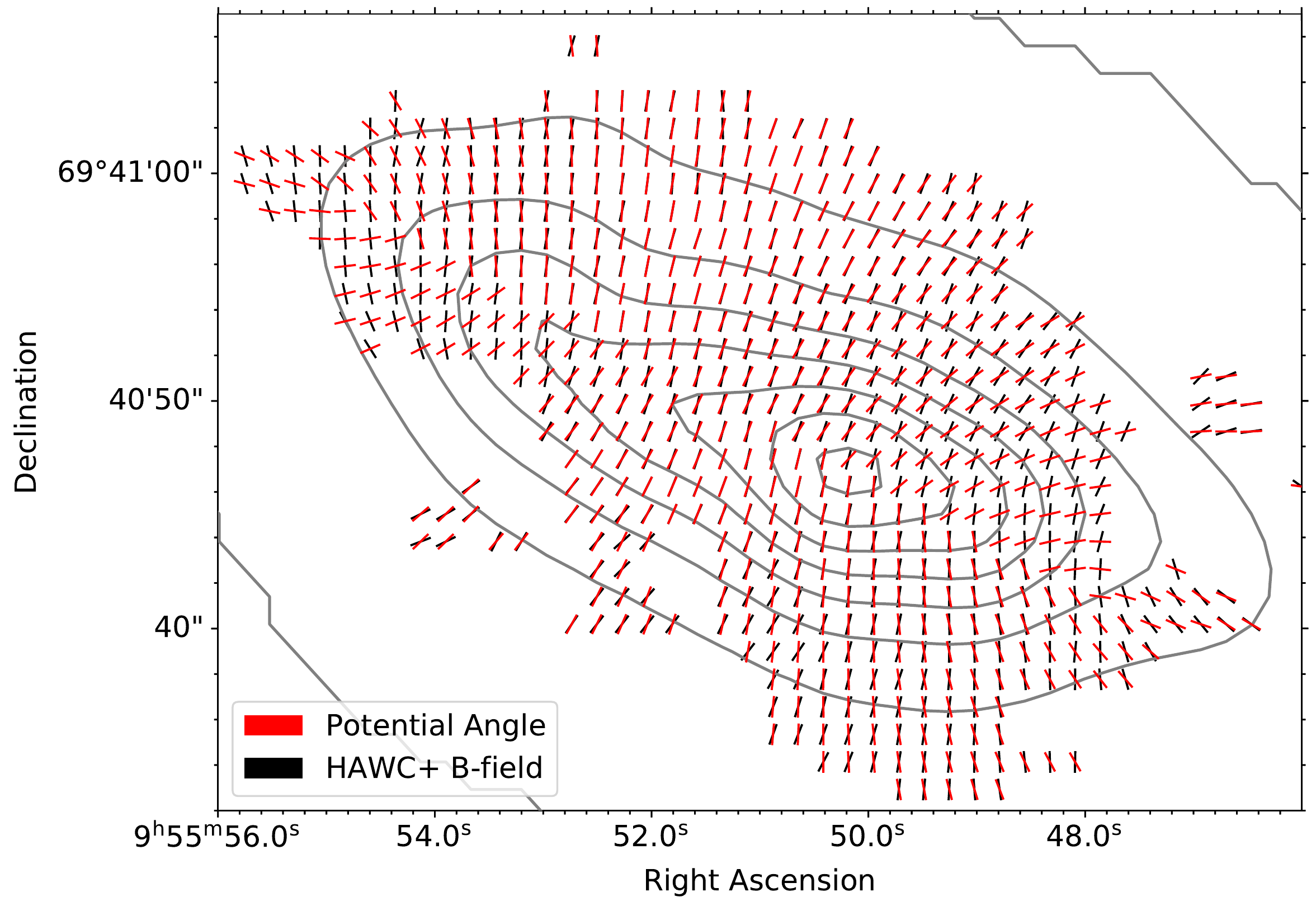}
\caption{Comparison between the HAWC+-inferred magnetic field orientation (black) and the potential orientation (red) in locations of M82 with $I>0.5$ Jy/arcsec$^{2}$ and $p > 3\sigma_{p}$. Above a below the center of the galaxy's plane (where the outflow is observed) both orientations coincide, hinting that the poloidal-type of magnetic field is near a force- free configuration. Near the FIR edges orientations differ suggesting a non-potential configuration of the field. Grey contours in the background correspond to M82's FIR intensity values from HAWC+.}
\label{fig:fig7}
\epsscale{2.}
\end{figure}

Using the values of $B_{\rm Norm}$ displayed in Figure \ref{fig:fig5} for the boundary condition $F(x)$, we can calculate the coefficients $a_n$ and $b_n$ in Eqs.~\ref{eq:coeff_a}, \ref{eq:coeff_b} for $n = 1 \dots 500$, $\ell = 1.2$ kpc (the extent of $F(x)$) using a trapezoidal method to evaluate the integral. Subsequently, we evaluate the potential field components in Eqs. \ref{eq:pot_Bx}, \ref{eq:pot_Bz} for $x,z =$ $-11.45$ to $11.45$ kpc. The resulting potential magnetic field lines are shown in Figure \ref{fig:fig6} within a $7.2 \times 7.2$ kpc$^{2}$ region centered at M82. Magnetic field lines are displayed in a grey scale that matches the potential field strength. Note that some lines appear truncated as an artifact of the line integration process performed by the python package \textsc{matplotlib} \citep{Hunter2007}. We display the potential field lines that originate from the empirical measurements using the $53$ \um~polarimetric observations with HAWC+. The maximum magnetic field strength of $\sim$1 mG seen in the bulk of the galaxy and the decrease to 100 $\mu$G at a radius of 1500 pc are in good agreement with the magnetic field strength estimated from HAWC$+$ data (Section \ref{subsection:PF_res}). 

Figure \ref{fig:fig7} displays a comparison between the orientations calculated with the potential field ($|\phi|$, Eq. \ref{eq:pot_ang}; red) and the measured with the HAWC+ instrument (black) for the central $\sim$ 700$\times$700 pc$^{2}$ region of M82 where the total intensity $I > 0.5$ Jy/arcec$^{2}$ and polarization fraction $P/\sigma_{P} > 3$ (the Chauvenet criteria). We measure an absolute angular difference (or misalignment) with a mean value of 16.3$\pm$25.6$^{\circ}$, where the maximum of occurrence is at $\sim$10.0$^{\circ}$. Smaller angular differences ($<$ 10$^{\circ}$) are located within the central $\sim$500 pc of the galaxy, above and below the plane. At these locations, the outflow dominates and the field orientation appears to be mainly poloidal. The strongest deviations are located on the far right and far left along the galactic plane. We find that the magnetic field is parallel to the galactic plane and outside the zone where the magnetic field is dragged by the superwind. \citet{Adebahr2017} also found a parallel magnetic field orientation in the western side of the galaxy (their labeled region 2), which is spatially coincident with ours. Early magnetic field models \citep[i.e.][]{Lesch1989} already accounted for these two components observed in radio observations \citep[i.e.][]{Reuter1992}. \citet{Beck2019} argued that dynamos are difficult to generate in starburst galaxies, as the dynamo growth rate is assumed to be shorter in starbursts than in regular galaxies. Thus, the magnetic field orientation along the galactic plane may be the remnant of the spiral magnetic field of M82 before the starburst stage.

\section{Discussion}\label{sec:DIS}

\subsection{The magnetic fields in the starburst region }\label{sec:PFvsB}

\begin{figure*}[ht!]
\includegraphics[angle=0,scale=0.36]{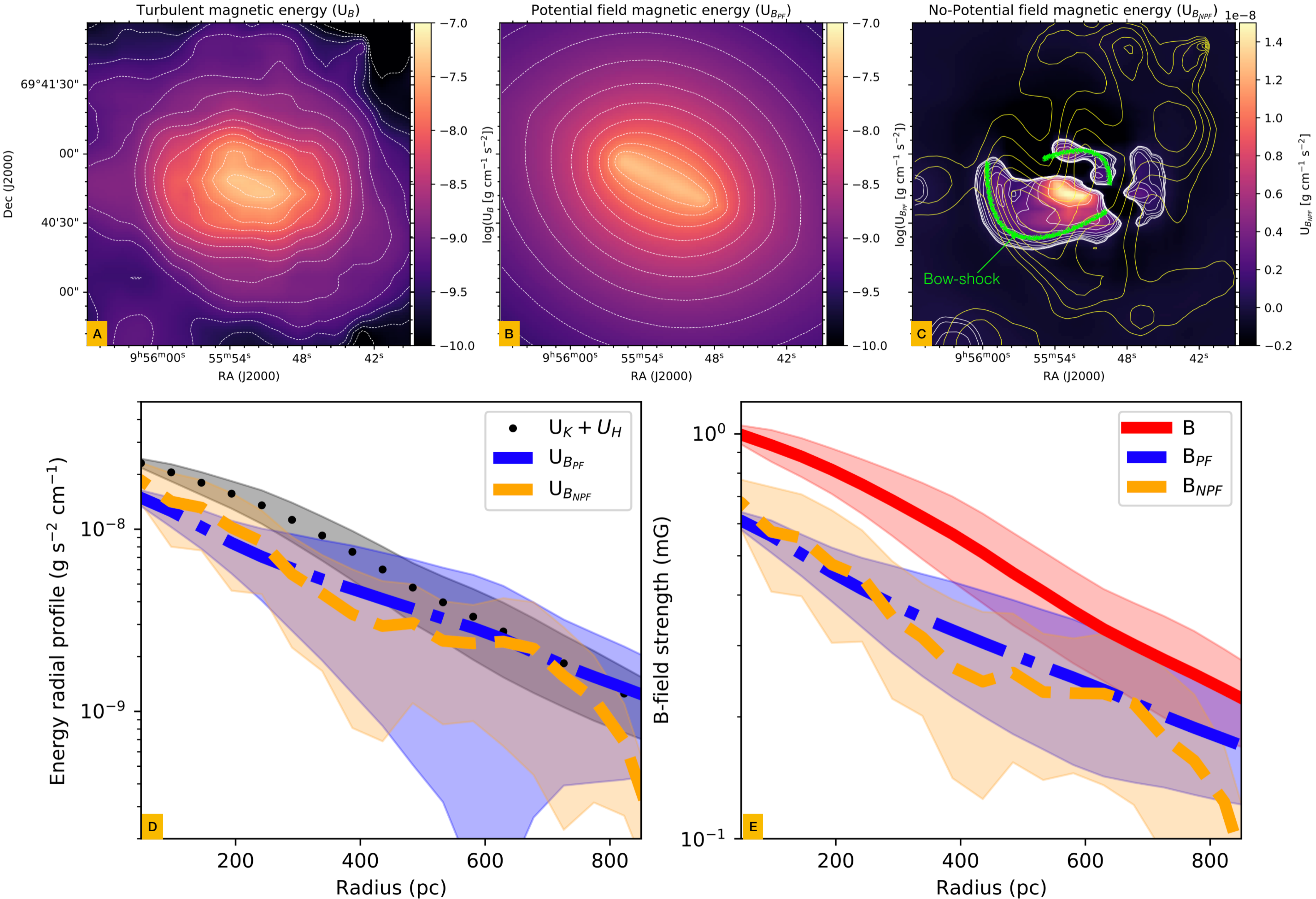}
\caption{The observed magnetic energy ($U_{B}$; A) is composed of a potential ($U_{B_{PF}}$; B) and a no-potential ($U_{B_{NPF}}$; C) field components. All energy maps with contours and FOV as Fig. \ref{fig:fig4}. The no-potential field magnetic energy, $U_{B_{NPF}} = U_{B}-U_{B_{PF}}$, with overlaid $^{12}CO(2-1)$ velocity dispersion (yellow contours) is shown. The contours start at $50$ km s$^{-1}$ and increases in steps of 5 km s$^{-1}$. The bow-shock-like pattern (green) is identified. 
(D) The radial profiles of $U_{K} + U_{H}$, $U_{B_{NPF}}$, and corrected potential field, $U_{B_{PF}}$, energies with their associated magnetic field strengths in (E). The median (lines) and 1$\sigma$ uncertainties are shown. \label{fig:fig8}}
\epsscale{2.}
\end{figure*}

As noted in Section \ref{sec:2dB}, the observed turbulent magnetic energy ($U_{B}$), is larger than the turbulent kinetic and hydrostatic energy ($U_{K} + U_{H}$) within the starburst mask (Fig. \ref{fig:fig4}). To investigate  additional components in the observed turbulent magnetic energy within the central $2.8\times2.8$ kpc$^{2}$ region, we subtract the potential field magnetic energy from the observed turbulent magnetic energy  (Fig. \ref{fig:fig8}). The resulting map is labeled `non-potential' field, $U_{B_{NPF}} = U_{B}-U_{B_{PF}}$, which can be interpreted as a `residual' map of the observed minus model magnetic field energies. The `non-potential' map, $U_{B_{NPF}}$, has a `bow-shock-like' arc structure along the south-east and north-west regions of the outflow with an extension up to $\sim600$ pc from the core (Fig. \ref{fig:fig8}-C). \citet{SBH1998} found a bow-shock arc at $\sim500$ pc in the south-east region of the outflow using [OIII] and H$_{\alpha}$ emission lines. We find that the morphology of the `non-potential' field is similar to the morphology of the $^{12}$CO(2-1) velocity dispersion. The increase of velocity dispersion corresponds to a higher `non-potential' magnetic energy. The south-east region has a higher contribution to the magnetic energy than the north-west region. The north-west region is viewed through the galactic disk, so this region of the outflow may be highly extinguished, make it appear less pronounced along the LOS. The bow-shock-like structure may be the wave front of the galactic outflow expanding through the intergalactic medium.  This result represents the first direct detection of the turbulent magnetic energy generated by a bow-shock in the galactic outflow of M82.

We conclude that the observed turbulent magnetic field energy within the starburst region is composed of two magnetic field components. A potential, large-scale (anisotropic turbulent or random) magnetic field from the galactic outflow, and a small-scale turbulent magnetic field (i.e. `non-potential' field)  from a bow-shock-like region. The large-scale (anisotropic turbulent or random) B-field is thought to be generated by the galactic wind at scales equal to or larger than the turbulent coherent length ($\delta = 73.6\pm5.6$ pc). The small-scale turbulent B-field is thought to be generated by the turbulent or random fields in the the bow-shock-like structure at scales smaller than $\delta$.

The observed turbulent magnetic energy has potential (large-scale) and non-potential (small-scale) components, $U_{B} = U_{B_{PF}} + U_{B_{NPF}}$. Fig. \ref{fig:fig8}-D shows the radial profiles of $U_{K}+U_{H}$, $U_{BPF}$, and $U_{B_{NPF}}$. Note that the non-potential magnetic energy (yellow) drops precipitously at a radius of $900$ pc. The magnetic fields associated to these energies are shown in Fig. \ref{fig:fig8}-E. We estimate that the non-potential and potential magnetic energies contribute $40\pm5$\% and $53\pm4$\% to the observed turbulent magnetic energy, respectively. Using these relative contributions, we estimate a median magnetic field strength of $305\pm15$ $\mu$G and $222\pm19$ $\mu$G for the potential field and non-potential magnetic field strength components, respectively.

Radio polarimetric observations using VLA and WSRT estimated a turbulent magnetic field strength in the range of $117-140$ $\mu$G in the central $1\times0.7$ kpc$^{2}$ of M82 \citep{Adebahr2017}. This magnetic field strength was estimated assuming equipartition between cosmic ray energies and magnetic field energies from \citet{Beck2005}. As the ionized gas is likely to be inhomogeneous in the starbursts region, \citet{Lucki2013,Lacki2013arxiv} accounted for the influence of giant star formation regions on the filing factor and radio luminosity. Specifically, uniform and discrete distribution of HII regions provides a high density and low filling factor medium, which will raise the average density and turbulent energy densities. Within the starburst volume, the filing factor is estimated to be in the range of $0.1-1$\% \citep[and references therein]{Lacki2013arxiv}, which may introduce an uncertainty in the estimation of the gas turbulent energies up to an order of magnitude. \citet{Lacki2013arxiv} suggested a magnetic field strength of $\sim300~\mu$G  for M82 assuming a supernova-driven turbulence outflow, which lead to a close equipartition between the ISM components within the starburst -- the turbulent energy density is comparable to the magnetic energy density in the starburst volume. To explain such a high magnetic field strength at radio wavelengths, \citet{Adebahr2013,Adebahr2017} suggested that there may be a superposition of at least two different phases of the magnetized medium at the core of M82 -- A strong mG field arising from the compact star-forming regions, and a weak diffuse $\mu$G component surrounding it. One of the main arguments for a two component field in M82 are the high synchrotron losses electrons would experience, which would lead to a non-visible radio halo at radio wavelengths \citep{Reuter1992,Adebahr2013}. This revision was suggested by \citet{Thompson2006}, who estimated a maximum B-field strength of $1.6$ mG for the starburst of M82 from an equipartition analysis between the magnetic energy density and the hydrostatic ISM pressure. \citet{Lacki2013} estimated a magnetic field strength of $220-240 \mu$G using a revised equipartition due to strong energy losses in the dense cores of starburst galaxies.

The two component scenario is similar to that measured in dense molecular clouds towards the Galactic center, where Zeeman splitting observations of OH masers indicate strong magnetic fields up to few mG \citep[e.g.][]{YZ1996}, while radio polarimetric observations reveal lower values of $50-100$ $\mu$G \citep[e.g.][]{Crocker2010}. Current FIR polarimetric observations at $53$ \um\ from HAWC+ of the Galactic center region using an approach similar to the one presented here estimate magnetic field strengths of $\sim$mG (Dowell et al., in preparation). From theoretical developments of the evolution of particles in the outflow of M82, \citet{YH2013} estimated a B-field strength in the range of $225-350$ $\mu$G, while \citet{paglione2012} found a best-fit model with magnetic field strengths of $450$ $\mu$G. The difference resides in the mean gas densities of $280-415$ cm$^{-3}$ for the former and $100-1000$ cm$^{-3}$ for the latter. Our results show that the mG component arises from the star forming region (potential field) and the small-scale turbulent field from the bow-shock-like feature (non-potential field) within the central $900$ pc of M82. Further models would require detail analysis of both components to explain the production of high energy particles, formation of galactic winds, and generation of galactic shocks.

\subsection{`Open' vs. `Closed' magnetic field lines}

We investigate whether the potential magnetic field lines of M82 are `open' (i.e. galactic outflows) or `closed' (i.e. galactic fountains). We have revised the definitions from solar physics \citep[i.e.][]{Levine1997,FS2001} to apply to  galactic winds. `Open' magnetic field lines remain attached to the central starburst and extend to large distances from the galactic plane. The turbulent kinetic energy of the outflowing wind exceeds the restoring turbulent magnetic energy. Field lines reaching these distances are dragged radially outward by the outflowing wind. `Open' field lines thus provide the missing link between the galaxies and intergalactic medium. `Closed' field lines remain attached to the central starburst and loop back to the plane of the galaxy. The turbulent magnetic energy exceeds the turbulent kinetic energy of the outflowing wind. `Closed' field lines provide a feedback channel from the central starburst back to the host galaxy.

We have shown that the turbulent kinetic and turbulent magnetic energies are in close equipartition up to $\sim2$ kpc (Fig. \ref{fig:fig4} and Section \ref{sec:2dB}). It is important to note that the field lines may have a complex morphology within the starburst region at higher angular resolution than those from our observations. The averaged orientation of the turbulent magnetic field appears as ordered and fairly parallel to the outflow lines, but there may be many reversals or loops at smaller scales. From observational results, however, the field lines appear to remain `open' up to $\sim2$ kpc.

The outflow of M82 extends at least $\sim11$ kpc from the galactic plane \citep[i.e.][]{DB1999}, at which distance some potential field lines may turn over and reconnect with the galactic plane. A magnetic field line will only serve as a channel for feedback if its strength is large enough to confine the ionized material to move along it. We find a striking similarity between the orientation of the observed B-field and potential field extrapolation, and the gas streams in MHD simulations using TNG50 \citep[see fig. 12 by][]{Nelson2019} or galactic outflows driven by supernovae. This result implies that the potential magnetic field is frozen in where the field lines are aligned with the outflowing wind. So do the magnetic field lines remain `open' at $\sim10$ kpc or do they turn over at a height above $\sim2$ kpc?

\begin{figure*}[ht!]
\centering
\includegraphics[angle=0,scale=0.7]{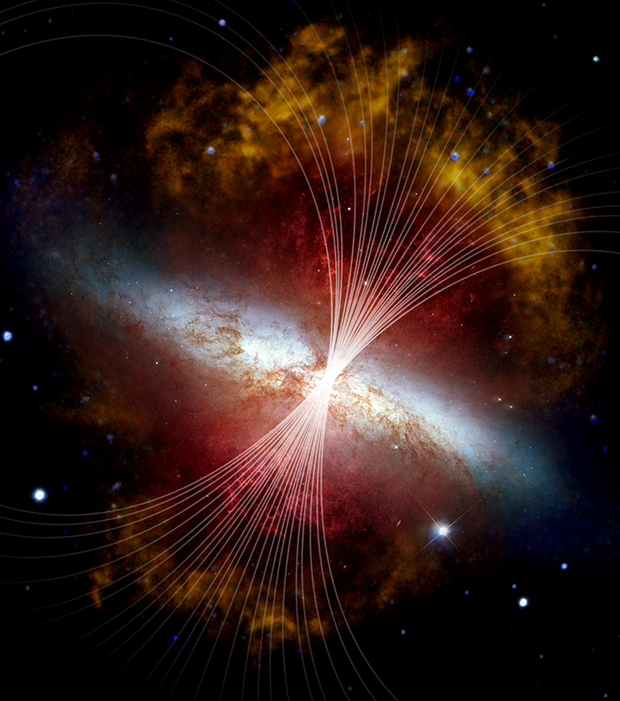}
\caption{Composite image of M82 displaying the potential magnetic field lines. The starburst activity is seen in white light and the high-speed galactic outflow (red) in H$\alpha$ emission, both by the \textit{HST}. In dark yellow, dust  observed with  \textit{Spitzer}. Potential field lines, in white, are seen vanished at $r\approx$ 7 kpc, as they are considered as `open' (galactic outflow).} 
\label{fig:fig9}
\epsscale{2.}
\end{figure*}

We can answer this question by comparing the kinetic and magnetic energies at such distances. To estimate the turbulent kinetic energy at distances of several kpc, the measurements of the dispersion velocity and mass density are required. HI observations of the M81 triplet covering an area of $3^{\circ}\times3^{\circ}$ at a resolution of $20\arcsec$ ($400$ pc) provides a velocity dispersion of $35\pm10$ km s$^{-1}$ at $6.6$ kpc from M82  \citep{deBlok2018}. \citet{Martini2018} estimated that dusty clouds in the outflow are embedded in an ambient medium. The cloud particle density is $n_{H}^{c} = 10$ cm$^{-3}$ ($\rho = 1.67 \times 10^{-23}$ g cm$^{-3}$), while the ambient particle density is $n_{H}^{a} = 0.044$ cm$^{-3}$ ($\rho = 7.36 \times 10^{-26}$ g cm$^{-3}$). Using Eq. \ref{eq:UK}, we estimate turbulent kinetic energies of $U_{K}^{c} = 3.07\times10^{-10}$  g s$^{-2}$ cm$^{-1}$ and  $U_{K}^{a} = 1.35\times10^{-12}$ g s$^{-2}$ cm$^{-1}$ for the clouds and ambient medium, respectively. 

Using the results of the potential field extrapolation, we estimate a magnetic field strength at radius $\ge6.6$ kpc of $B_{PF} \le15$ $\mu$G, yielding $U_{B_{PF}} \le  8.9 \times 10^{-12}$  g s$^{-2}$ cm$^{-1}$. Magnetic field strengths up to $50$ $\mu$G at scales of $\sim8$ kpc in M82 has been measured using radio polarimetric observations and justified in terms of magnetized galactic winds \citep[i.e.][]{Kronberg1999}. \citet{Basu2013} used radio polarimetric observations to estimate B-field strengths of $\sim10~\mu$G at a distance of $\sim10$ kpc in several nearby normal spiral galaxies assuming equipartition of energy between cosmic ray particles and magnetic fields. We estimate that $U_{B_{PF}} \sim U_{K}^{a} < U_{K}^{c}$ in the outflow at scales up to $6.6$ kpc. We find that the dusty clouds are dominated by the kinetic energy in the outflowing wind, while the ambient medium is in close equipartition with the magnetic energy. Since the turbulent kinetic energy dominate the dusty medium, we conclude that the field lines are `open' at distances up to $6.6$ kpc from the plane of M82, channeling magnetic energy and matter into the intergalactic medium.

Magnetic field strengths in the range of $2-40$ $\mu$G have been measured in clusters on scales of $3-10$ kpc through Faraday rotation measurements \citep[i.e.][]{Dreher1987,Kim1990,Taylor1993,Clarke2001}. These magnetic fields may be primordial, seeded in the intergalactic medium from shock waves or linked with the formation and evolution of galaxies \citep[i.e.]{Subramanian2019}. Using semi-analytic simulations of magnetized galactic winds, \citet{Bertone2006, Sumui2018} suggested that galactic outflows may be able to seed a fraction of the magnetic field in the intracluster medium.

\section{Conclusions}\label{sec:CON}

We used the HAWC+ polarimetric data as well as median values of the mass density, $\rho$, and the velocity dispersion, $\sigma_{v,^{12}\rm CO(2-1)}$, from the literature to estimate the average plane-of-the-sky magnetic field strength in the starburst region of M82 to be $B = 1.04\pm0.17$ mG (DCF and angular dispersion method).

We described a novel approach to quantify the turbulent magnetic field when large-scale flows are present (Section \ref{subsec:Boutflow}). We modified the DCF method to account for galactic superwind by adding a steady-flow term to the wave equation, which reverts to the traditional approach when large-scale flows are negligible. After we accounted for the large-scale flow, the median magnetic field within the starburst region is reduced to $B = 0.77\pm0.17$ mG.

We defined the turbulent plasma beta, $\beta'$, as the ratio of hydrostatic-plus-turbulent-kinetic pressure to magnetic pressure and, using median values within the starburst mask with a size of $873\times510$ pc$^{2}$ from Section \ref{sec:Bfield}, estimate $\beta' = 0.56\pm0.23$. The turbulent magnetic field energy is larger than the turbulent kinetic energy within the starburst. We can then use the pixel-by-pixel values of the density and velocity dispersion to construct, for the first time, a two-dimensional map of the turbulent magnetic field strength within the central $2.8\times 2.8$ kpc$^{2}$ region of M82 (Figure \ref{fig:fig4}-D).

We modified the solar potential field method to work with galactic outflows using HAWC+ polarization data. We extrapolated the magnetic field from the core using the Laplace equation 
and investigate the potential magnetic structures along the galactic outflow of M82. 
The resulting potential magnetic field structure is shown in Figure \ref{fig:fig6}. These results indicate that the observed turbulent magnetic field energy within the starburst region is composed of two components: a large-scale (anisotropic turbulent) field arising from the galactic outflow and a small-scale turbulent field arising from a bow-shock-like region. This result represents the first detection of the magnetic energy from a bow shock in the galactic outflow of M82.

The results of the potential field extrapolation allow us to determine, for the first time, if the field lines are `open' (i.e. galactic outflow) or `closed' (i.e. galactic fountain). We show that the turbulent kinetic and turbulent magnetic energies are in close equipartition up to $\sim2$ kpc (measured), while the turbulent kinetic energy dominates at $\sim7$ kpc (extrapolated). We estimated a magnetic field strength $\le15~\mu$G at distances $\ge6.6$ kpc from the starburst region. We conclude that the magnetic field lines associated with the galactic superwind of M82 are `open', channeling matter into the galactic halo and beyond. These observations indicate that the powerful winds associated with the starburst phenomenon could be responsible for injecting material enriched with elements like carbon and oxygen into the intergalactic medium.

We demonstrated that FIR polarization observations are a powerful tool to study the B-field morphology in the cold and dense ISM of galactic outflows. Ongoing efforts like the SOFIA Legacy Program (PIs: Lopez-Rodriguez \& Mao) focused on studying extragalactic magnetism will provide deeper observations at $53$ \um\ to analyze the large-scale magnetic field in the disk of M82 as well as other nearby galaxies. The results presented here can also be used to investigate the high energy particles production from starburst galaxies. This work serves as a strong reminder of the potential importance of magnetic fields, often completely overlooked, in the formation and evolution of galaxies.


\acknowledgments

Based on observations made with the NASA/DLR Stratospheric Observatory for Infrared Astronomy (SOFIA). SOFIA is jointly operated by the Universities Space Research Association, Inc. (USRA), under NASA contract NAS2-97001, and the Deutsches SOFIA Institut (DSI) under DLR contract 50 OK 0901 to the University of Stuttgart. 

%

\vspace{5mm}
\facilities{SOFIA (HAWC+), \textit{Herschel} (PACS, SPIRE), \textit{HST} (WFPC2, ACS)}


\software{\textsc{astropy} \citep{astropy2013,astropy2018}, 
\textsc{APLpy} \citep{RB2012},
\textsc{matplotlib} \citep{Hunter2007},
\textsc{Python} \citep{py3},
\textsc{Numpy} \citep{numpy},
\textsc{pandas} \citep{pandas}. 
          }

\end{document}